\documentstyle[epsfig]{mn2e}

\title[Strange quark stars]{Dynamical stability of strange quark stars}

\author[P. S. Negi]{P. S. Negi$^{1}$\thanks{E-mail:
negi@aries.ernet.in; psnegi\_nainital@yahoo.com} \\
$^{1}$ Department of Physics, Kumaun University,
              Nainital - 263 002, India \\
}
\date{Accepted ------ .
      Received ------ ;
      }
\pagerange{\pageref{firstpage}--\pageref{lastpage}} 
\pubyear{2006}

\begin{document}

\maketitle

\label{firstpage}

\begin{abstract}
  The necessary and sufficient condition for dynamical stability is worked out for the sequences of relativistic
  star models corresponding to the well defined and causal values of adiabatic sound speed,$\sqrt ({\rm d}P/{\rm d}E)_0
  = v_0$, at the centre. On the basis of these conditions, we show that the mass-radius 
  $(M-R)$ relation corresponding to the MIT bag models of strange quark matter (SQM) and
  the models obtained by Day et al (1998) do not provide the necessary and sufficient condition 
  for dynamical stability for the equilibrium configurations, since such configurations can 
  not even fulfill the necessary condition of hydrostatic equilibrium provided by the 
  exterior Schwarzschild solution. These findings will remain unaltered and can be extended
  to any other sequence of pure SQM. This study explicitly show that although the strange quark matter
  might exist in the state of zero pressure and temperature, but the models of pure strange quark
  `stars' can not exist in the state of hydrostatic equilibrium on the basis of General Relativity Theory.
  This study can affect the results which are claiming that various objects like - RX J1856.5-3754,
  SAX J1808.4-3658, 4U 1728-34, PSR 0943+10 etc. might be strange stars.

\end{abstract}


\section{Introduction}

         A lot of astrophysical interest has emerged in the strange star models of SQM 
         following the X-ray observations of compact sources like RX J1856.5-3754, 
         SAX J1808.4-3658, 4U 1728-34 and PSR 0943+10 etc. by the satellite {\em Chandra} and {\em XMM-Newton}.
In particular, the strange star models of Day et al (1998) have been presented as the most successful
description of these sources (Drake et al 2002; Li et al 1999a; Li et al 1999b; Xu 2002). Some authors,
however, have shown that the MIT bag models of SQM can also explain the small value of the radiation radius,
$R_\infty$, ranging from 3.8 - 8.2 km (Drake et al 2002) for a mass of about 1$M_\odot$
corresponding to the compact star RX J1856.5-3754 (Kohri et al 2003; but for an alternative model,
see, e.g. Walter \& Lattimer 2002).

Although the possibility of the existence of hypothetical self-bound SQM stars dates back to eighties (Witten 1984; Fahri \& Jaffe 1984;
Haensel et al 1986; Alcock et al 1986 etc.), but the possibility of the existence of such stars
made of pure SQM (i.e. corresponding to a single equation of state (EOS) of self-bound matter)
in the state of hydrostatic equilibrium have never been questioned. Recently, Negi \& Durgapal
 (2001) have obtained a `compatibility criterion' for ascertaining the state of hydrostatic equilibrium
 in static spherical structures. This criterion states that ``for each and every assigned value of $\sigma [\equiv (P_0/E_0) \equiv $ the 
ratio  of central  pressure  to   central   energy-density], the compactness 
ratio $u (\equiv M/R)$ of the entire configuration should not  exceed the 
compactness ratio, $u_h$, of the corresponding homogeneous density sphere 
(that is, $u \leq  u_h$)''. On the basis of this criterion, they have demonstrated that the `regular' configurations
corresponding to a single EOS or density variation do not exist in the state of hydrostatic equilibrium.
As an example, they have shown this inconsistency, particularly for the stiffest EOS of self-bound matter, 
$P = (E - E_s)$ (where $P$ is the pressure, $E$ is the energy-density and $E_s$ represents the  surface 
density at zero  pressure; we consider $G = c = 1$ (i.e., the geometrized  units)). The reason
 for this inconsistency
is explained by Negi (2004a; 2006). From this explanation, one can easily extract the following
statement that `regular' configurations
corresponding to a single EOS or density variation {\em can} not exist in the state of hydrostatic equilibrium.
Even more recently, the `compatibility criterion' is established
as Theorem 1  and the dependency of necessary and sufficient condition for dynamical
stability provided by the mass-radius ($M-R$) relation on this theorem
is obtained as Theorem 2 (and its subsequent corollaries) in Negi (2007; 2008).

The present study deals with the construction of such theorems for different values of the parameter
$a$ in the range $0.301 \leq a \leq 0.463$ corresponding to the EOS $P = a (E - E_s)$.
This range can cover the MIT bag models of SQM (called the SQM0, SQM1 and SQM2 in the literature;
see, e.g. Haensel et al 2006, and references therein) and the models of Day et al (1998) (called SS1 and SS2)
mentioned above. While the MIT bag models, we consider in the present study, provide the masses of strange stars as large as those `observed' for any neutron star, the sequences of Day et al (1998) can produce the most compact configurations of strange stars among various models of SQM available in the literature. These sequences are, therefore, widely used in the literature to explain: the relatively smaller
values of the stellar radius `observed' for the sources like RX J1856.5-3754 (Drake et al 2002), SAX J1808.4-3658 (Li et al 1999a)
and the kHz quasi-periodic oscillations (QPOs) observed among various low-mass X-ray binaries (LMXBs) (see, e.g., Li et al 1999b; Zdunik 2000; and references therein).

 %__________________________________________________________________

\section{The necessary and sufficient condition for dynamical stability for strange quark star models}

     The absolute upper bound on compactness ratio for any dynamically stable regular configuration
     may be obtained by using the `compressible' sphere of homogeneous energy-density (Negi 2004b), since
     the following relation holds good for a constant $\Gamma_1$

$$ \frac{\Gamma_1P}{P + E} = \frac{{\rm d}P}{{\rm d}E}. $$

Obviously, the adiabatic speed of sound, $\sqrt({\rm d}P/{\rm d}E) \equiv v$, will become finite 
inside such configuration for a finite (constant) value of $\Gamma_1$. In order to have a 
desired value of $({\rm d}P/{\rm d}E)$ at the centre of `compressible' homogeneous density sphere,
one can work out a particular value of $u$ and the corresponding (critical) value of constant 
$\Gamma_1$ for which the configuration remains pulsationally stable. This particular value of
$u$ represents an absolute upper bound, $u_{\rm max,abs}$, consistent with that of the condition
$ v^2 \leq ({\rm d}P/{\rm d}E)$ and dynamical stability, since it follows from the `compatibility criterion'
(Theorem 1 of Negi 2007; and references therein) that corresponding to this particular $u$ value $(=u_{\rm max,abs})$, any regular
configuration can not have a value of the ratio of central pressure to central energy-density, $P_0/E_0$
(or, equivalently, a central value of the `local' adiabatic index $(\Gamma_1)_0 (\equiv [(P_0 + E_0)/P_0](dP/dE)_0)$
less than (greater than) that of the homogeneous density sphere. Notice that this result may be generalized for a number of relativistic 
star sequences provided that every member of that particular sequence satisfies the condition
$({\rm d}P/{\rm d}E)_0 \equiv v_0^2 = a$ (say) (here and elsewhere in this paper, the subscript ``0''
refers the value of the corresponding quantity at the centre). It follows from the previous studies (Negi 2007; 2004b)
that in order to assure the necessary and sufficient condition of dynamical stability for a mass, the
maximum stable value of $u (= u_{\rm max})$(corresponding to the first maximum along masses in
the stable branch of mass-radius ($M-R$) relation) and the corresponding central value of the `local'
adiabatic index $(\Gamma_1)_0$ of the said sequence must satisfy the inequalities $u_{\rm max} \leq u_{\rm max,abs}$
and $(\Gamma_1)_0 \leq (\Gamma_1)_{\rm 0,max,abs}$ ($\equiv $(critical) constant $\Gamma_1$) simultaneously.

Following the previous study (Negi 2004b), we obtain various $u$ and the (critical) constant $\Gamma_1$
values for dynamically stable configuration corresponding to various values of $({\rm d}P/{\rm d}E)_0(\equiv v_0^2 = a)$
at the centre of `compressible' homogeneous density sphere as shown in Table 1. Since these results are also applicable to
the sequences of strange quark stars characterized by the pure EOS $P = a(E - E_s)$, their applicability
(according to various values of the parameter $a$ appearing in this EOS as shown in column 5) is indicated in column 1
of Table 1. In view of the discussion of last paragraph, the various $u$ and the (critical)
constant $\Gamma_1$ values are indicated here as $u_{\rm max,abs}$ and $(\Gamma_1)_{\rm 0,max,abs}$
respectively.

\begin{table}
\begin{center}
      \caption[]{The absolute maximum stable values of compactness ratio ($u_{\rm max,abs} \equiv M/R$)  
      and the corresponding central value of the `local' adiabatic index $(\Gamma_1)_{\rm 0,max,abs}$ for the equilibrium
      sequences satisfying the condition ${\rm d}P/{\rm d}E = a \equiv v^2$ at the centre of the configuration as shown in column 5.
      These values of the parameter $a$ which are appearing in the EOS, $P = a(E - E_s)$, correspond to
      the models of strange quark stars shown in column 1.
      A comparison of the pair of these absolute values  with those of the pair of 
      corresponding values obtained for the models of pure strange quark matter (Tables 3 - 7) 
      clearly shows that the mass-radius diagram corresponding to any sequences of the strange quark stars (shown in Fig.1) 
      does not provide
      a necessary and sufficient condition for dynamical stability for the equilibrium configurations.}

%\vspace{1.0cm}

\begin{tabular}{ccccc}

\hline
${\rm Model}$ & $u_{\rm max,abs}$ & $(\Gamma_1)_{\rm 0,max,abs}$ & ${(P_0 / E_0)}$ & $a(\equiv v_0^2)$ \\
\hline
{\rm SQM1} & 0.224446 & 1.734653 & 0.209953 & 0.301000 \\
{\rm SQM2} & 0.232527 & 1.764514 & 0.224920 & 0.324000 \\
{\rm SQM0} & 0.235633 & 1.776597 & 0.230958 & 0.333333 \\
{\rm SS2}  & 0.268911 & 1.932300 & 0.307994 & 0.455000 \\
{\rm SS1}  & 0.270718 & 1.942420 & 0.312960 & 0.463000 \\
\hline

\end{tabular}
\end{center}
\end{table}

%______________________________________________________________

\section{Compatibility criterion and the sequences of strange quark stars}
The metric for spherically symmetric and static configurations can be written
in the following form (remembering that we are using geometrized units, i.e. $G = c = 1$; where $G$ and $c$ represent respectively, the universal constant of gravitation and the speed of light in vacuum) 
\begin{equation}
ds^2 =  e^{\nu} dt^2 - e^{\lambda} dr^2 - r^2 d\theta^2 - r^2 $ sin$^2 \theta d\phi^2 , 
\end{equation}
where $\nu$ and $\lambda$ are functions of $r$ alone.  The  Oppenheimer-Volkoff 
(O-V) equations (Oppenheimer \& Volkoff 1939), resulting  from  Einstein's field  equations, for 
systems with isotropic pressure $P$  and  energy-density  $E$  can  be 
written as

\begin{figure}
   \centering
%   \vspace{270pt}
%   \includegraphics[height=9.5cm,width=9.5cm]{mr.ps}
\epsfig{file=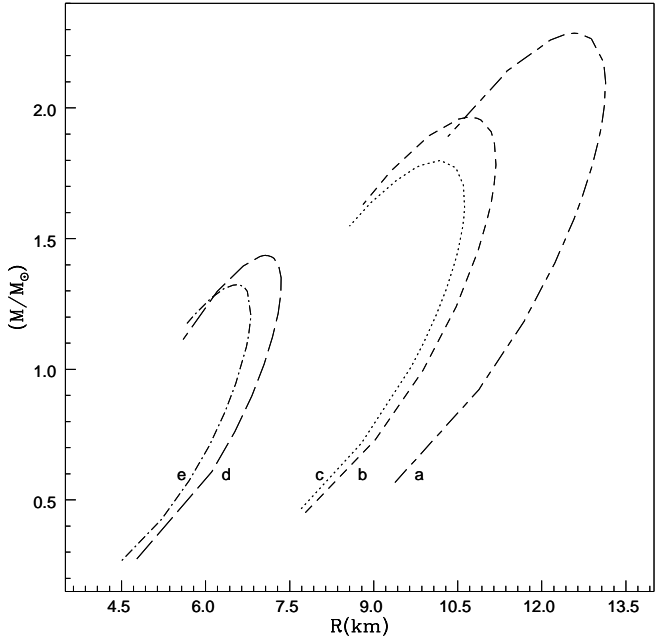,height=9.5cm,width=9.5cm} 
      \caption{The mass-radius diagram for the models corresponding to the pure
       EOS, $P = a(E - E_s)$, for the well defined value of the parameters $a$ and $E_s$
       shown in Table 2. The labels `a' - `e' (from right to left) represent the models SQM2, 
       SQM0, SQM1, SS1 and SS2 respectively.
        The models do not fulfill the `compatibility criterion'
        as shown in Tables 3 - 7. Also, the mass-radius diagram does not provide the 
        necessary and sufficient condition for dynamical stability for any of the sequence 
        because the inequalities, $u_{\rm max} \leq u_{\rm max,abs}$ and $(\Gamma_1)_0 \leq (\Gamma_1)_{\rm 0,max,abs}$,
        are not fulfilled simultaneously at the maximum value of mass for any of the sequence
         as shown in Tables 3 - 7.
              }
         \label{FigVibStab}
   \end{figure}

\begin{eqnarray}
P' & = & - (P + E)[4 \pi P r^3 + m]/r(r - 2m) \\                       
\nu'/2 & = & - P'/(P + E)  \\                                            
m'(r) & = & 4\pi E r^2 \,;
\end{eqnarray}
where $ m(r) = \int_{0}^{r} 4\pi Er^2 dr $ is the mass, contained within the  radius  $r$,  
and  the prime denotes radial derivative.

In order to solve equations (2) - (4), we rewrite the linear EOS mentioned in the last section
\begin{equation}
P = a(E - E_s).
\end{equation}
Since, the various SQM models considered in the present study (corresponding to different values of
the parameters $a$ and $E_s$ shown in Table 2) are very precisely approximated by this EOS (Zdunik 2000; Gondek-Rosi\'{n}ska et al 2000).
Moreover, this form also turns out to be exact for massless quarks (free or interacting) corresponding
to a value of $a = 1/3$. The first three pairs of the parameters $a$ and $E_s$ shown in Table 2 represent
the MIT bag models of SQM corresponding to: the mass of a strange quark $m_s = 200$ MeV, a value of QCD
coupling constant $\alpha_c = 0.2$ and a bag constant B = 56MeVfm$^{-3}$ (indicated as SQM1 
in Table 2); $m_s = 100$ MeV, $\alpha_c = 0.6$ and B = 40MeVfm$^{-3}$ (indicated as SQM2 in Table 2)
; $m_s = 0$,\,$\alpha_c = 0$ and B = 60MeVfm$^{-3}$ (indicated as SQM0 in Table 2). The last two entries
(indicated as SS2 and SS1 in Table 2), however, represent the SQM models of Day et al (1998) corresponding to
density dependent quark masses and a color dependent vector interquark potential.

Equation (5) is solved together with the coupled equations (2) - (4) for five pairs of the parameters
$a$ and $E_s$ (shown in Table 2) until the pressure vanishes at the surface of the configuration.

\begin{table}
\begin{center}
      \caption[]{The values of the parameters $a$ and $E_s$ appearing in the EOS, $P = a(E - E_s)$ 
      for the models of strange quark stars shown in column 1.}

%\vspace{1.0cm}

\begin{tabular}{ccc}

\hline
${\rm Model}$ & $a(\equiv v^2)$ & $E_s (10^{14}$gcm$^{-3}$) \\
\hline
{\rm SQM1} & 0.301000 & 4.50000 \\
{\rm SQM2} & 0.324000 & 3.05600 \\
{\rm SQM0} & 0.333333 & 4.27850 \\
{\rm SS2}  & 0.455000 & 13.3200 \\
{\rm SS1}  & 0.463000 & 11.5300 \\
\hline

\end{tabular}
\end{center}
\end{table}

At the surface, $r = R$, we obtain 
\begin{equation}
 P = 0,\, E = E_s,\, m(r = R)  =  M, 
\end{equation}
and
\begin{equation}
e^{\nu} = e^{-\lambda} = (1 - 2M/R) = (1 - 2u). 
\end{equation}
The results of the calculations are shown in Table 3 - 7 and the $M-R$ 
diagram is presented in Fig.1. 

It follows from Table 3 that along the stable branch of the SQM0 
sequence, the maximum value of mass ($M_{\rm max} \simeq 1.9654M_\odot$) corresponds to the maximum `stable' value of $u_{\rm max} 
\simeq 0.2707$ and the corresponding `local' value of $(\Gamma_1)_0 \simeq 1.5974$. Although, 
this value of $(\Gamma_1)_0$ turns out to be consistent with that of the absolute upper bound 
on $(\Gamma_1)_0 ((\Gamma_1)_{0,\rm max, abs} \simeq 1.7766)$, the maximum `stable' value of
 $u_{\rm max} \simeq 0.2707$ is found to be inconsistent with that of the absolute upper 
 bound on $u_{\rm max} (u_{\rm max,abs} \simeq 0.2356)$. Thus the configuration turns out to 
 be inconsistent with that of the findings of the last section. 
 
It is seen from Table 4 (which presents the results for SQM1 sequence) that along the stable branch of the 
sequence, the maximum value of mass ($M_{\rm max} \simeq 1.7996M_\odot$) corresponds to the maximum `stable' value of $u_{\rm max} 
\simeq 0.2608$ and the corresponding `local' value of $(\Gamma_1)_0 \simeq 1.5531$. The pair of these values,
however, show inconsistency with that of the pair of absolute upper bounds 
     $u_{\rm max,abs} (\simeq 0.2244)$ and $(\Gamma_1)_{0,\rm max,abs} (\simeq 1.7346)$. It follows, therefore, that 
 the $M-R$ relation corresponding to the SQM1 EOS does not provide a necessary and 
 sufficient condition for dynamical stability for the equilibrium configurations.
 
 The results of the calculations for SQM2 sequence is presented in Table 5. It follows from this table
 that along the stable branch of the 
sequence, the maximum value of mass ($M_{\rm max} \simeq 2.2859M_\odot$) corresponds to the maximum `stable' value of $u_{\rm max} 
\simeq 0.2684$ and the corresponding `local' value of $(\Gamma_1)_0 \simeq 1.5798$. The pair of these values
is also found to be inconsistent with that of the pair of absolute upper bounds 
     $u_{\rm max,abs} (\simeq 0.2325)$ and $(\Gamma_1)_{0,\rm max,abs} (\simeq 1.7645)$ obtained for the SQM2
     sequence in the last section. It follows, therefore, that 
 the $M-R$ relation corresponding to the SQM2 EOS does not provide a necessary and 
 sufficient condition for dynamical stability for the equilibrium configurations.
 
The maximum value of mass ($M_{\rm max} \simeq 1.4358M_\odot$) models yields the maximum value of 
   $u (u_{\rm max}) \simeq 0.2999$ and the corresponding value of $(\Gamma_1)_0 \simeq 1.7793$
    as shown in Table 6. The pair of these values, however, also show inconsistency with that of the pair of 
    absolute upper bounds, $u_{\rm max,abs}(\simeq 0.2707)$ and $(\Gamma_1)_{0,\rm max,abs} 
    (\simeq 1.9424)$. It follows from this result that the necessary and 
    sufficient condition for dynamical stability (provided by the $M-R$ relation) remains 
    unsatisfied even for the SS1 models.
    
 Finally, it follows from Table 7 that along the stable 
   branch of the sequence in the mass-radius relation for the SS2 sequence, the maximum value of mass
   , $M_{\rm max} \simeq 1.3241M_\odot$,  corresponds to the maximum `stable' value of $u_{\rm max} 
\simeq 0.2980$ and the corresponding value of $(\Gamma_1)_0 \simeq 1.7738$. The pair of these values
is also found to be inconsistent with that of the pair of absolute upper bounds 
     $u_{\rm max,abs} (\simeq 0.2689)$ and $(\Gamma_1)_{0,\rm max,abs} (\simeq 1.9323)$ 
     obtained for the SS2
     sequence in the last section. It follows, therefore, that 
 the $M-R$ relation corresponding to the SS2 EOS does not provide a necessary and 
 sufficient condition for dynamical stability for the equilibrium configurations. 
   
It follows from the above findings that none of
 the $M-R$ relation corresponding to the SQM sequences considered in the present study
 (SQM0, SQM1, SQM2, SS1 and SS2) provide the necessary and 
 sufficient condition for dynamical stability for the equilibrium configurations. As the 
 total mass `$M$' that appears in the exterior Schwarzschild solution (equations 6 - 7) does 
 not 
 fulfill the definition of the `actual mass' 
 that should be present in the exterior Schwarzschild solution (Negi 2004a; 2006), the 
 equilibrium sequences corresponding to SQM0, SQM1, SQM2, SS1 and SS2 EOS do not, therefore,
  even fulfill the 
 necessary condition of hydrostatic equilibrium (Negi 2007), as a result the 
 `compatibility criterion' can not be satisfied by such configurations. This is also evident 
 from the comparison of column 2 and 6 of Tables 3 - 7 that for each assigned value of $(P_0/E_0),
 \, u > u_h$.

\section{Results and conclusions}
We have obtained the necessary and sufficient condition for dynamical stability applicable 
to the
MIT bag models of strange quark matter as well the models of Day et al (1998). Our 
investigation show that the mass-radius relation corresponding to the
MIT bag models of strange quark matter (called SQM0, SQM1 and SQM2 in the literature) and those of the models of Day et al (1998)
(called SS1 and SS2) do not provide the necessary and sufficient condition for dynamical 
stability because at the maximum value of mass along the stable branch of the mass-radius relation, the pair of 
the maximum `stable' value of compactness, $u_{\rm max}$, and the corresponding central value
 of the `local' adiabatic index, $(\Gamma_1)_0$, turns out to be inconsistent with that of 
 the pair of {\em absolute} values:
$u_{max,abs}$ and $(\Gamma_1)_{0,{\rm max,abs}}$, compatible with the structure of 
general relativity, dynamical stability and the condition, 
$({\rm d}P/{\rm d}E)_0 \leq v_0^2 = a$. The reason behind this inconsistency lies in the fact 
that the `compatibility criterion' (Negi \& Durgapal 2001) can not be fulfilled by any 
of the sequence, composed of regular configurations corresponding to a {\em single} EOS 
with finite (non-zero) values of surface and central density (Negi 2004a; 2006) like - the EOS (equation 5) of self-bound strange quark matter. Since this form (equation 5) turns out to be independent of the strange quark matter model (Haensel et al 2006), the conclusions of this study
regarding the dynamical stability and hydrostatic equilibrium will remain valid for any model of strange quark matter. Following the present study, the values $u_{max,abs}$ and $(\Gamma_1)_{0,{\rm max,abs}}$, however, may be obtained for any set of the parameters $a$ and $E_s$.

The present study does not exclude the construction of a two-density (core-envelope) SQM model. One can always construct such a model corresponding to a core
defined by a SQM EOS and an envelope (significantly thick crust) defined by some other suitable EOS
so that the necessary and sufficient condition for dynamical stability could be satisfied for the resulting sequence. The satisfaction of the last condition would autometically guarantee the `appropriate' fulfillment of the `compatibility criterion' (see, e.g. Negi 2008). However, if the core of such a two-density model is described by any of the SQM model considered in the present study, the upper bound on the compactness of stable configuration can not exceed the values shown in Table 1, which are significantly less than those of the values shown in Tables 3 - 7 for the pure SQM models.

\section*{Acknowledgments}

The author gratefully acknowledges the Aryabhatta Research Institute of Observational Sciences (ARIES), 
Nainital for providing library and computer-centre facilities.

%\newpage

\begin{table*}
\begin{center}
      \caption[]{Mass ($M$), size ($R$), compactness ratio ($u \equiv M/R$)  
      and the central value of the `local' adiabatic index $(\Gamma_1)_0$ of the  
      configuration for various values of the ratio of central pressure to central 
      energy-density ($P_0/E_0$) as  obtained  by substituting the well defined values of 
      the parameter $a = (1/3)$ and  
      the surface density $E_s  = 4.2785 \times 10^{14}$ g\, cm$^{-3}$ (so called the SQM0 model) in the 
EOS, $P = a(E - E_s)$. It is seen that for each  assigned  value  of $P_0/E_0$, the inequality, $u \leq
u_h$  (where $u_h$  represents the corresponding value of the  compactness ratio 
for the homogeneous density distribution shown in column 2) always remains unsatisfied.  The 
values in italics correspond to the limiting case upto which the configuration 
remains pulsationally `stable'. Also, the $M-R$ relation does not provide the necessary and
sufficient condition for dynamical stability, as the sequence does not satisfy the inequalities,
namely $u_{\rm max} \leq 0.2356$ and $(\Gamma_1)_0 \leq 1.7766$ simultaneously at the maximum value of mass.}

\begin{tabular}{ccccccc}

\hline
${(P_0 / E_0)}$ & $u_h$ & $(\Gamma_1)_0$ & $(M/M_{\odot})$ & $R({\rm km})$ & $u$ & $z_R$ \\
\hline
0.05000 & 0.08318 & 7.00000 & 0.45082 & 7.78008 & 0.08558 & 0.09842 \\
0.07500 & 0.11496 & 4.77778 & 0.73037 & 9.02544 & 0.11952 & 0.14636 \\
0.10000 & 0.14202 & 3.66667 & 0.99678 & 9.87869 & 0.14903 & 0.19358 \\
0.12516 & 0.16543 & 2.99659 & 1.24019 & 10.4701 & 0.17495 & 0.24025 \\
0.15000 & 0.18550 & 2.55556 & 1.45289 & 10.8610 & 0.19758 & 0.28582 \\
0.16570 & 0.19686 & 2.34500 & 1.57148 & 11.0243 & 0.21054 & 0.31429 \\
0.17517 & 0.20328 & 2.23625 & 1.63663 & 11.0945 & 0.21788 & 0.33128 \\
0.18680 & 0.21076 & 2.11777 & 1.70876 & 11.1506 & 0.22634 & 0.35170 \\
0.20000 & 0.21875 & 2.00000 & 1.78215 & 11.1793 & 0.23546 & 0.37479 \\
0.21760 & 0.22864 & 1.86520 & 1.86325 & 11.1566 & 0.24667 & 0.40489 \\
0.23045 & 0.23538 & 1.77978 & 1.90955 & 11.0944 & 0.25422 & 0.42630 \\
0.25000 & 0.24490 & 1.66667 & 1.95416 & 10.9166 & 0.26439 & 0.45677 \\
0.25570 & 0.24752 & 1.63694 & 1.96079 & 10.8435 & 0.26708 & 0.46515 \\
0.26200 & 0.25036 & 1.60560 & 1.96454 & 10.7510 & 0.26989 & 0.47408 \\
0.26370 & 0.25110 & {\sl 1.59740} & {\sl 1.96538} & {\sl 10.7238} & {\sl 0.27069} & {\sl 0.47665} \\
0.26750 & 0.25276 & 1.57944 & 1.96382 & 10.6580 & 0.27215 & 0.48135 \\
0.27500 & 0.25596 & 1.54545 & 1.95718 & 10.5121 & 0.27499 & 0.49069 \\
0.29500 & 0.26402 & 1.46328 & 1.89320 & 9.98117 & 0.28015 & 0.50808 \\
0.31150 & 0.27019 & 1.40342 & 1.75850 & 9.30441 & 0.27915 & 0.50464 \\
0.32002 & 0.27323 & 1.37494 & 1.63092 & 8.80823 & 0.27348 & 0.48570 \\
\hline

\end{tabular}
\end{center}
\end{table*}

%\newpage
\begin{table*}
\begin{center}
\caption[]{Mass ($M$), size ($R$), compactness ratio ($u \equiv M/R$)  
      and the central value of the `local' adiabatic index $(\Gamma_1)_0$ of the  
      configuration for various values of the ratio of central pressure to central 
      energy-density ($P_0/E_0$) as  obtained  by substituting the well defined values of the
      parameter $a = 0.301$ and 
      the surface density $E_s  = 4.5 \times 10^{14}$ g\, cm$^{-3}$ (so called the SQM1 model) in  the 
EOS, $P = a(E - E_s)$. It is seen that for each  assigned  value  of $P_0/E_0$, the inequality, $u \leq
u_h$  (where $u_h$  represents the corresponding value of the  compactness ratio 
for the homogeneous density distribution shown in column 2) always remains unsatisfied.  The 
values in italics correspond to the limiting case upto which the configuration 
remains pulsationally `stable'. Also, the $M-R$ relation does not provide the necessary and
sufficient condition for dynamical stability, as the sequence does not satisfy the inequalities,
namely $u_{\rm max} \leq 0.2244$ and $(\Gamma_1)_0 \leq 1.7346$ simultaneously at the maximum value of mass.}

%\vspace{1.0cm}

\begin{tabular}{ccccccc}

\hline
${(P_0 / E_0)}$ & $u_h$ & $(\Gamma_1)_0$ & $(M/M_{\odot})$ & $R({\rm km})$ & $u$ & $z_R$ \\
\hline
0.05250 & 0.08660 & 6.03433 & 0.46646 & 7.70766 & 0.08939 & 0.10349 \\
0.07560 & 0.11566 & 4.28248 & 0.71805 & 8.78498 & 0.12072 & 0.14817 \\
0.10390 & 0.14587 & 3.19802 & 1.00885 & 9.66885 & 0.15411 & 0.20231 \\
0.11360 & 0.15510 & 2.95065 & 1.10139 & 9.89138 & 0.16446 & 0.22072 \\
0.12533 & 0.16558 & 2.70266 & 1.20749 & 10.1157 & 0.17631 & 0.24285 \\
0.13720 & 0.17550 & 2.49488 & 1.30755 & 10.2964 & 0.18757 & 0.26504 \\
0.14950 & 0.18512 & 2.31438 & 1.40417 & 10.4408 & 0.19864 & 0.28808 \\
0.15570 & 0.18973 & 2.23420 & 1.44924 & 10.4968 & 0.20392 & 0.29952 \\
0.16560 & 0.19679 & 2.11863 & 1.51712 & 10.5657 & 0.21208 & 0.31780 \\
0.17022 & 0.19997 & 2.06930 & 1.54734 & 10.5899 & 0.21581 & 0.32642 \\
0.17661 & 0.20423 & 2.00532 & 1.58555 & 10.6116 & 0.22069 & 0.33795 \\
0.18270 & 0.20818 & 1.94851 & 1.62008 & 10.6232 & 0.22525 & 0.34901 \\
0.20000 & 0.21875 & 1.80600 & 1.70414 & 10.6015 & 0.23742 & 0.37992 \\
0.22000 & 0.22994 & 1.66918 & 1.77124 & 10.4680 & 0.24992 & 0.41397 \\
0.24040 & 0.24033 & {\sl 1.55308} & {\sl 1.79962} & {\sl 10.19029} & {\sl 0.26084} & {\sl 0.44591} \\
0.25650 & 0.24789 & 1.47449 & 1.78033 & 9.83403 & 0.26739 & 0.46613 \\
0.26000 & 0.24947 & 1.45869 & 1.77007 & 9.73554 & 0.26854 & 0.46977 \\
0.27000 & 0.25384 & 1.41582 & 1.72204 & 9.39867 & 0.27062 & 0.47641 \\
0.28000 & 0.25804 & 1.37600 & 1.63684 & 8.94980 & 0.27013 & 0.47484 \\
0.28650 & 0.26067 & 1.35161 & 1.54863 & 8.56832 & 0.26695 & 0.46475 \\
\hline

\end{tabular}
\end{center}
\end{table*}

\newpage

%\newpage

\begin{table*}
\begin{center}
\caption[]{Mass ($M$), size ($R$), compactness ratio ($u \equiv M/R$)  
      and the central value of the `local' adiabatic index $(\Gamma_1)_0$ of the  
      configuration for various values of the ratio of central pressure to central 
      energy-density ($P_0/E_0$) as  obtained  by substituting the well defined values of the
      parameter $a = 0.324$ and 
      the surface density $E_s  = 3.056 \times 10^{14}$ g\, cm$^{-3}$ (so called the SQM2 model) in  the 
EOS, $P = a(E - E_s)$. It is seen that for each  assigned  value  of $P_0/E_0$, the inequality, $u \leq
u_h$  (where $u_h$  represents the corresponding value of the  compactness ratio 
for the homogeneous density distribution shown in column 2) always remains unsatisfied.  The 
values in italics correspond to the limiting case upto which the configuration 
remains pulsationally `stable'. Also, the $M-R$ relation does not provide the necessary and
sufficient condition for dynamical stability, as the sequence does not satisfy the inequalities,
namely $u_{\rm max} \leq 0.2325$ and $(\Gamma_1)_0 \leq 1.7645$ simultaneously at the maximum value of mass.}

%\vspace{1.0cm}

\begin{tabular}{ccccccc}

\hline
${(P_0 / E_0)}$ & $u_h$ & $(\Gamma_1)_0$ & $(M/M_{\odot})$ & $R({\rm km})$ & $u$ & $z_R$ \\
\hline
0.05250 & 0.08660 & 6.49543 & 0.56708 & 9.37718 & 0.08932 & 0.10340 \\
0.07950 & 0.12014 & 4.39947 & 0.92222 & 10.8749 & 0.12525 & 0.15509 \\
0.10050 & 0.14251 & 3.54788 & 1.18523 & 11.6854 & 0.14981 & 0.19490 \\
0.11960 & 0.16055 & 3.03303 & 1.40659 & 12.2302 & 0.16987 & 0.23067 \\
0.13539 & 0.17403 & 2.71709 & 1.57458 & 12.5696 & 0.18502 & 0.25992 \\
0.14480 & 0.18152 & 2.56157 & 1.66776 & 12.7300 & 0.19350 & 0.27724 \\
0.15740 & 0.19097 & 2.38245 & 1.78416 & 12.9010 & 0.20426 & 0.30026 \\
0.16724 & 0.19793 & 2.26134 & 1.86790 & 13.0011 & 0.21220 & 0.31808 \\
0.17550 & 0.20350 & 2.17015 & 1.93345 & 13.0637 & 0.21860 & 0.33300 \\
0.18472 & 0.20946 & 2.07801 & 2.00122 & 13.1113 & 0.22544 & 0.34948 \\
0.20000 & 0.21875 & 1.94400 & 2.09921 & 13.1365 & 0.23602 & 0.37627 \\
0.21466 & 0.22705 & 1.83336 & 2.17693 & 13.0991 & 0.24546 & 0.40155 \\
0.24000 & 0.24013 & 1.67400 & 2.26543 & 12.8779 & 0.25983 & 0.44286 \\
0.25344 & 0.24649 & 1.60241 & 2.28480 & 12.6679 & 0.26639 & 0.46300 \\
0.25800 & 0.24857 & {\sl 1.57981} & {\sl 2.28593} & {\sl 12.57937} & {\sl 0.26840} & {\sl 0.46932} \\
0.26050 & 0.24969 & 1.56776 & 2.28507 & 12.52605 & 0.26944 & 0.47264 \\
0.26500 & 0.25168 & 1.54664 & 2.28147 & 12.42296 & 0.27125 & 0.47844 \\
0.27500 & 0.25596 & 1.50218 & 2.25958 & 12.15167 & 0.27465 & 0.48954 \\
0.29500 & 0.26402 & 1.42230 & 2.14040 & 11.37537 & 0.27791 & 0.50046 \\
0.31150 & 0.27019 & 1.36413 & 1.89047 & 10.32081 & 0.27054 & 0.47616 \\
\hline

\end{tabular}
\end{center}
\end{table*}

\begin{table*}
\begin{center}
      \caption[]{Mass ($M$), size ($R$), compactness ratio ($u \equiv M/R$)  
      and the central value of the `local' adiabatic index $(\Gamma_1)_0$ of the  
      configuration for various values of the ratio of central pressure to central 
      energy-density ($P_0/E_0$) as  obtained  by substituting the well defined values of the
      parameter $a = 0.463$ and 
      the surface density $E_s  = 1.153 \times 10^{15}$ g\, cm$^{-3}$ (so called the SS1 model) in  the 
EOS, $P = a(E - E_s)$. It is seen that for each  assigned  value  of $P_0/E_0$, the inequality, $u \leq
u_h$  (where $u_h$  represents the corresponding value of the  compactness ratio 
for the homogeneous density distribution shown in column 2) always remains unsatisfied.  The 
values in italics correspond to the limiting case upto which the configuration 
remains pulsationally `stable'. Also, the $M-R$ relation does not provide the necessary and
sufficient condition for dynamical stability, as the sequence does not satisfy the inequalities,
namely $u_{\rm max} \leq 0.2707$ and $(\Gamma_1)_0 \leq 1.9424$ simultaneously at the maximum value of mass.}

%\vspace{1.0cm}

\begin{tabular}{ccccccc}

\hline
${(P_0 / E_0)}$ & $u_h$ & $(\Gamma_1)_0$ & $(M/M_{\odot})$ & $R({\rm km})$ & $u$ & $z_R$ \\
\hline
0.05000 & 0.08318 & 9.72300 & 0.27461 & 4.77066 & 0.08502 & 0.09767 \\
0.10000 & 0.14202 & 5.09300 & 0.60812 & 6.11130 & 0.14697 & 0.19009 \\
0.12530 & 0.16555 & 4.15813 & 0.76088 & 6.52334 & 0.17228 & 0.23518 \\
0.15060 & 0.18595 & 3.53737 & 0.89857 & 6.82630 & 0.19442 & 0.27916 \\
0.17554 & 0.20353 & 3.10058 & 1.01937 & 7.04448 & 0.21373 & 0.32159 \\
0.20000 & 0.21875 & 2.77800 & 1.12303 & 7.19491 & 0.23054 & 0.36219 \\
0.22494 & 0.23254 & 2.52133 & 1.21316 & 7.29239 & 0.24571 & 0.40224 \\
0.25000 & 0.24490 & 2.31500 & 1.28875 & 7.34108 & 0.25929 & 0.44125 \\
0.27500 & 0.25596 & 2.14664 & 1.34992 & 7.34561 & 0.27143 & 0.47903 \\
0.30000 & 0.26593 & 2.00633 & 1.39499 & 7.30456 & 0.28207 & 0.51470 \\
0.32500 & 0.27496 & 1.88762 & 1.42418 & 7.21845 & 0.29141 & 0.54823 \\
0.33333 & 0.27778 & 1.85201 & 1.43026 & 7.17907 & 0.29426 & 0.55892 \\
0.34500 & 0.28159 & 1.80503 & 1.43479 & 7.11390 & 0.29789 & 0.57287 \\
0.34900 & 0.28286 & 1.78965 & 1.43523 & 7.08850 & 0.29905 & 0.57740 \\
0.35173 & 0.28371 & {\sl 1.77934} & {\sl 1.43584} & {\sl 7.07090} & {\sl 0.29992} & {\sl 0.58084} \\
0.35750 & 0.28548 & 1.75810 & 1.43452 & 7.02958 & 0.30141 & 0.58674 \\
0.36500 & 0.28775 & 1.73149 & 1.43203 & 6.97206 & 0.30337 & 0.59463 \\
0.37500 & 0.29066 & 1.69767 & 1.42483 & 6.88509 & 0.30566 & 0.60398 \\
0.39500 & 0.29620 & 1.63515 & 1.39556 & 6.67082 & 0.30899 & 0.61794 \\
0.42500 & 0.30383 & 1.55241 & 1.29832 & 6.20827 & 0.30888 & 0.61746 \\
\hline

\end{tabular}
\end{center}
\end{table*}

\begin{table*}
\begin{center}
      \caption[]{Mass ($M$), size ($R$), compactness ratio ($u \equiv M/R$)  
      and the central value of the `local' adiabatic index $(\Gamma_1)_0$ of the  
      configuration for various values of the ratio of central pressure to central 
      energy-density ($P_0/E_0$) as  obtained  by substituting the well defined values of the
      parameter $a = 0.455$ and 
      the surface density $E_s  = 1.332 \times 10^{15}$ g\, cm$^{-3}$ (so called the SS2 model) in  the 
EOS, $P = a(E - E_s)$. It is seen that for each  assigned  value  of $P_0/E_0$, the inequality, $u \leq
u_h$  (where $u_h$  represents the corresponding value of the  compactness ratio 
for the homogeneous density distribution shown in column 2) always remains unsatisfied.  The 
values in italics correspond to the limiting case upto which the configuration 
remains pulsationally `stable'. Also, the $M-R$ relation does not provide the necessary and
sufficient condition for dynamical stability, as the sequence does not satisfy the inequalities,
namely $u_{\rm max} \leq 0.2689$ and $(\Gamma_1)_0 \leq 1.9323$ simultaneously at the maximum value of mass.}

%\vspace{1.0cm}

\begin{tabular}{ccccccc}

\hline
${(P_0 / E_0)}$ & $u_h$ & $(\Gamma_1)_0$ & $(M/M_{\odot})$ & $R({\rm km})$ & $u$ & $z_R$ \\
\hline
0.05200 & 0.08592 & 9.20500 & 0.26824 & 4.50668 & 0.08791 & 0.10151 \\
0.07700 & 0.11728 & 6.36409 & 0.42638 & 5.21413 & 0.12078 & 0.14826 \\
0.10123 & 0.14324 & 4.94972 & 0.57332 & 5.70461 & 0.14844 & 0.19258 \\
0.12551 & 0.16573 & 4.08021 & 0.70879 & 6.06609 & 0.17258 & 0.23575 \\
0.15054 & 0.18590 & 3.47745 & 0.83583 & 6.34366 & 0.19461 & 0.27954 \\
0.17500 & 0.20318 & 3.05500 & 0.94515 & 6.53925 & 0.21348 & 0.32101 \\
0.21230 & 0.22575 & 2.59819 & 1.08665 & 6.72921 & 0.23851 & 0.38280 \\
0.22560 & 0.23288 & 2.47184 & 1.12924 & 6.76898 & 0.24640 & 0.40414 \\
0.25310 & 0.24634 & 2.25271 & 1.20426 & 6.80984 & 0.26119 & 0.44698 \\
0.30422 & 0.26752 & 1.95063 & 1.29776 & 6.74808 & 0.28405 & 0.52162 \\
0.32162 & 0.27379 & 1.86971 & 1.31466 & 6.68467 & 0.29048 & 0.54480 \\
0.33333 & 0.27778 & 1.82001 & 1.32098 & 6.62845 & 0.29435 & 0.55927 \\
0.33512 & 0.27837 & 1.81272 & 1.32200 & 6.61925 & 0.29499 & 0.56168 \\
0.34500 & 0.28159 & {\sl 1.77384} & {\sl 1.32409} & {\sl 6.56203} & {\sl 0.29803} & {\sl 0.57341} \\
0.34750 & 0.28238 & 1.76435 &  1.32350 & 6.54519 & 0.29866 & 0.57589 \\
0.35520 & 0.28478 & 1.73597 & 1.32204 & 6.49229 & 0.30076 & 0.58417 \\
0.37500 & 0.29066 & 1.66833 & 1.30782 & 6.32604 & 0.30535 & 0.60271 \\
0.39530 & 0.29628 & 1.60602 & 1.27268 & 6.10024 & 0.30814 & 0.61435 \\
0.41500 & 0.30137 & 1.55139 & 1.21008 & 5.80516 & 0.30788 & 0.61324 \\
0.42240 & 0.30320 & 1.53218 & 1.17546 & 5.66673 & 0.30638 & 0.60697 \\
\hline

\end{tabular}
\end{center}
\end{table*}

\end{document}